\documentclass[english,aps,
onecolumn,
nofootinbib]{revtex4}

\usepackage[T1]{fontenc}
\usepackage[latin9]{inputenc}
\usepackage{amsmath}

\makeatletter
\@ifundefined{textcolor}{}
{%
 \definecolor{BLACK}{gray}{0}
 \definecolor{WHITE}{gray}{1}
 \definecolor{RED}{rgb}{1,0,0}
 \definecolor{GREEN}{rgb}{0,1,0}
 \definecolor{BLUE}{rgb}{0,0,1}
 \definecolor{CYAN}{cmyk}{1,0,0,0}
 \definecolor{MAGENTA}{cmyk}{0,1,0,0}
 \definecolor{YELLOW}{cmyk}{0,0,1,0}
 }

\makeatother

\usepackage{babel}

\usepackage{graphicx}
\usepackage[T1]{fontenc}
\usepackage[latin9]{inputenc}

\newcommand{\be}{\begin{equation}}
\newcommand{\ee}{\end{equation}}
\newcommand{\bea}{\begin{eqnarray}}
\newcommand{\eea}{\end{eqnarray}}

\newcommand\dalonk{\dot{A}_{\lon k}}
\newcommand\ddalonk{\ddot{A}_{\lon k}}
\newcommand\daperk{\dot{A}_{\bot k}}
\newcommand\ddaperk{\ddot{A}_{\bot k}}

\newcommand{\gapp}{\mathrel{\raise.3ex\hbox{$>$}\mkern-14mu
              \lower0.6ex\hbox{$\sim$}}}
\newcommand{\lapp}{\mathrel{\raise.3ex\hbox{$<$}\mkern-1mu
              \lower0.6ex\hbox{$\sim$}}}

\newcommand\gsim{\gtrsim}

\newcommand\vev[1]{{\langle {#1} \rangle}}
\renewcommand\({\left(}
\renewcommand\){\right)}

\newcommand\eq[1]{Eq.~(\ref{#1})}
\newcommand\eqs[2]{Eqs.~(\ref{#1}) and (\ref{#2})}

\newcommand\eqst[2]{Eqs.~(\ref{#1})--(\ref{#2})}

\newcommand\eqreff[1]{(\ref{#1})}

\newcommand\pa{\partial}

\newcommand\mpl{M_{\rm P}}

\newcommand{\dlabel}[1]{\label{#1}}

\def\call{{\cal L}}

\def\calp{{\cal P}}

\def\calpz{{\calp_\zeta}}

\newcommand\bfk{{\mathbf k}}

\newcommand\bfx{{\mathbf x}}

\newcommand\eV{\,\mbox{eV}}

\newcommand\sub[1]{_{\rm #1}}
\newcommand\su[1]{^{\rm #1}}

\newcommand\mtwo{^{-2}}
\newcommand\mthree{^{-3}}

\newcommand\mn{{\mu\nu}}

\newcommand\bfkp{{{\bfk}'}}

\newcommand\aper{A_{\bot\bfk}}
\newcommand\aperk{A_{\bot k}}
\newcommand\lon{{||}}
\newcommand\alon{A_{\lon\bfk}}
\newcommand\alonk{A_{\lon k}}

\begin{document}

\title{On the health of a vector field  with $RA^2/6$ coupling to gravity}

\author{Mindaugas Kar\v{c}iauskas}
\email{m.karciauskas@lancaster.ac.uk}
\affiliation{Department of Physics, Lancaster University, Lancaster LA1 4YB, UK\
}
\author{David H. Lyth}
\email{d.lyth@lancaster.ac.uk}
\affiliation{Department of Physics, Lancaster University, Lancaster LA1 4YB, UK\
}

\begin{abstract}
The coupling $RA^2/6$ of a vector field to gravity was proposed as a mechanism
for generating a primordial magnetic field, and more recently 
as a mechanism for generating a statistically anisotropic contribution to
the primordial curvature perturbation. In either case, the vector field's
perturbation has both a transverse and a longitudinal component,
and the latter has some unusual features which call into question the health
of the theory. We calculate for the first time the energy density generated
by the longitudinal field perturbations, and go on to argue that the theory
may well be healthy in at least some versions.
\end{abstract}

\maketitle

\section{Introduction}

As far as we can tell, the initial condition for the formation of structure in the universe is provided by the primordial curvature perturbation $\zeta$, which is known to exist when cosmological scales start to come inside the horizon. In turn, $\zeta$ is supposed ultimately to originate during an early era of inflation, as the vacuum fluctuation of one or more bosonic fields. Until recently these were always taken  to be scalar fields, but following the proposal of Dimopoulos \cite{kostas}  vector fields are receiving considerable attention \cite{sy,dklr,more1,KDandMK,more2,nonabelian,vecinf,vecinf2,peloso1,peloso2}. If a vector field contributes, statistical anisotropy of $\zeta$ may be expected, providing a smoking gun whose form will be a powerful discriminator between different  models for generating the curvature perturbation.

In order to give a significant contribution to the curvature perturbation over a wide range of scales, the spectrum of the perturbation in each relevant field should be nearly  scale-invariant. Barring an unlikely cancellation, this requires inflation to be almost exponential (de Sitter) while cosmological scales are leaving the horizon. Near scale-invariance is then automatic for scalar fields that are light (mass much less than the Hubble parameter)  with canonically kinetic terms and minimal coupling to gravity. For a vector field in contrast, those same requirements lead to a contribution to the spectrum $\calpz$ going like $k^3$. This almost certainly makes the contribution  negligible on cosmological scales \cite{dklr}, because $\calpz$ is constrained to be at most of order 1 on the much smaller scale leaving the horizon at the end of inflation (to avoid  excessive black hole production).

For a light vector field $A_\mu$ with the canonical kinetic term, one recovers scale invariance by introducing a non-minimal coupling  $R A^2/6$ to gravity \cite{KDandMK,dklr}. The spectrum of each transverse mode is   then the same as for a scalar field, while the spectrum of the longitudinal mode is twice as big. The latter feature generates statistical anisotropy of a distinctive form \cite{dklr,more1}.

A coupling $R A^2/6$ has also been invoked in a completely different context,
namely the case when $A_\mu$ describes the electromagnetic field \cite{turnerwidrow}. In that
case,  the perturbation generated during inflation becomes a primordial magnetic field
which may be cosmologically significant. 

The coupling $R A^2/6$ thus has at least two possible uses, but concerns
have been raised about its health. They are about 
 the longitudinal mode \cite{peloso1,dklr,peloso2} and are of two kinds. 
The first concern is that  the effective kinetic term of the longitudinal mode
(taking into account the non-minimal coupling) is negative on sub-horizon scales during 
inflation \cite{peloso1,dklr}.
 As a result, 
one suspects that the corresponding particles carry negative energy density, allowing them
to be created from the vacuum which makes it unstable. 
Such creation is known to occur for a 
minimally coupled scalar field with a negative-sign canonical kinetic term; a scalar field with
this property is called a ghost and is cosmologically unacceptable. 

The second concern \cite{peloso2} applies only if the action of $A_\mu$ has a non-zero mass term.
Taking the inflationary expansion to be isotropic, the longitudinal  mode function
 becomes singular at a certain epoch after inflation is over.
Taking into account the anisotropy in the inflationary expansion, caused at some level
by the  homogeneous part of $A_\mu$, 
further singularities occur at least in the evolution equations
\cite{peloso2}.

If $A_\mu$ describes the electromagnetic field the mass term is absent and there are no
singularities.
If instead $A_\mu$ is  supposed to generate a contribution to the 
curvature perturbation we cannot simply drop  the mass term, but we will see in Section
\ref{sfive} how it might be replaced by something 
more complicated so as to avoid the singularities. 
If the singularities do exist they
 indicate a breakdown of the linear evolution, and progress in understanding
what is going on could be made only with a non-linear calculation which is not performed here.  
In the absence of such a calculation, we are free to assume
that any  singularities  have a negligible 
effect; in other words, that the linearly-evolved quantities
match before and after an epoch of singularity. 

In this paper, we focus mostly on the possibility of generating a contribution to
the curvature perturbation. In Section \ref{stwo} we review known results.
In Section \ref{sthree} we give the contribution of the vacuum fluctuation to the
energy density, and in Section \ref{sfour} we give the contribution of particles along
with a discussion of the issue of particle creation from the vacuum.
In Section \ref{sfive} we mention ways of avoiding a mass term in the
action of the field and we conclude in Section \ref{ssix}.

\section{Brief review of known results}

\dlabel{stwo}

\subsection{Action}

Except where stated the notation and conventions of \cite{dklr} are adopted. We are considering the action

\bea
S &=& \int d\eta d^3x \sqrt{-g} \call \nonumber \\
\call &=&  \frac12 m_P^2 R
-\frac14 F_{\mu\nu}F^{\mu\nu}-\frac12 M^2 A^2 \label{L} \\
M^2 &\equiv&    m^2 + \frac16 R,
\dlabel{eq:Mass-effective}
\eea
where $F_\mn \equiv \pa_\mu A_\nu-\pa_\nu A_\mu$ and we are ignoring the back-reaction, which means that the line element is   

\be
ds^2= -dt^2 + a^2(t) \delta_{ij} x_i x_j = a^2(\eta) \( -1 + \delta_{ij} x_i x_j \).
\dlabel{rw} \ee
The curvature scalar is

\begin{equation} 
R=-6\left(\dot{H}+2H^{2}\right)=3\left(3P/\rho-1\right)H^{2},
\dlabel{eq:Ricci-scl}
\end{equation}
where $P$ and $\rho$ are the pressure and energy density of the dominant component of the universe. During exponential inflation $R=-12H^2$. Going from the coordinate-induced basis to an orthogonal basis we arrive at the physical
vector field $B_\mu= A_\mu/a$. (This notation for  $B_\mu$ and $A_\mu$ is the reverse of
the one in  \cite{dklr}.) 

This action is supposed to hold during inflation. It was first invoked \cite{turnerwidrow}
to generate a primordial magnetic field, with $m=0$ and $A_\mu$ the electromagnetic 
vector potential. Then it was invoked in \cite{KDandMK,dklr} to allow
 $A_\mu$  to generate a contribution to the curvature perturbation,
which might be done through the curvaton mechanism \cite{curvaton} or one of its
generalisations \cite{newbook}. 

The kinetic term of the action is the only gauge-invariant expression (confining ourselves
as usual to a term that is quadratic in spacetime  derivatives of $A_\mu$).
Additional possibilities \cite{dklr} 
exist if the kinetic term breaks gauge invariance. One 
 might consider them on the ground that
 gauge invariance is broken by the coupling $RA^2/6$ and (if it exists)
the mass term. But they would not give the desired flat spectrum for $\delta A$
and we therefore reject them in the present context.\footnote{In the context of generating
a magnetic field one might want to consider them, because a flat spectrum for
$\delta A$ is then not particularly desirable as we see after \eq{mageq}.}

\subsection{Transverse and longitudinal modes}

We write the vector field as a sum of the homogeneous part ant the perturbation, $A_\mu(\bfx,t)=A_\mu(t) + \delta A_\mu(\bfx,t)$. The time component of the homogeneous part $A_\mu(t)$ vanishes, and the physical space components satisfy
 
\be
\ddot B_i(t) + 3H(t) \dot B_i(t) + m^2 B_i(t) =0,
\dlabel{unpert} \ee
which is the same as for a scalar field with the mass-squared  $m^2$. While the time component of the perturbation is related to the space components $\delta A_i$ by a constraint equation. 

In what follows we work with the Fourier components of $\delta A_i(t,\bfx)$ defined by

\be
A_{i\bfk}(t) = \int d^3x \delta A_i(t,\bfx) e^{-i\bfk\cdot\bfx}
. \ee
(In some formulas we use the physical momentum $p\equiv k/a$.)$A_{i\bfk}(t)$ can be written as 

\begin{equation}
A_{i\bfk}(t) =\sum_{\lambda}e_i^{\lambda}(\bfk)
A_{\lambda\bfk}(t),
\label{eq:pol-vecs-sep}\end{equation}
where $e_i^{\lambda}$ are polarization vectors. With the $z$ axis chosen to be along the $\mathbf{k}$ direction,
the polarization vectors are

\begin{equation}
e^{\mathrm{L}}\equiv\frac{1}{\sqrt{2}}\left(1,i,0\right),\quad e^{\mathrm{R}}
\equiv\frac{1}{\sqrt{2}}\left(1,-i,0\right)\quad\mathrm{and}\quad 
e^{||}\equiv\left(0,0,1\right),
\label{eq:pol-vecs-def}\end{equation}
and we choose $e_i^*(\bfk)= e_i(-\bfk)$ so that  $A^*_{\lambda\bfk}(t) = A_{\lambda,-\bfk}(t)$ making $A_\lambda(t,\bfx)$ real. 
(In \cite{dklr} both of these relations had a minus sign.) The  transverse components $L$ and $R$ satisfy the same equations and we will use $\lambda =\bot$ to denote either of them.

The vector field is quantized by promoting it to an operator in the Heisenberg picture:
\bea
\hat A_{\lambda\bfk}(t) &=& \hat a_\lambda(\bfk) A_{\lambda k}(t) +
 \hat a^\dagger_\lambda(-\bfk) A^*_{\lambda k}(t)  \\
\left[\hat{a}_{\lambda}\left(\mathbf{k}\right),\hat{a}_{\lambda'}^{\dagger}
\left(\mathbf{k}'\right)\right] &=& \left(2\pi\right)^{3}\delta
\left(\mathbf{k}-\mathbf{k}'\right)\delta_{\lambda\lambda'}
\eea
Around the epoch of horizon exit corresponding to $aH=k$, the vacuum fluctuation of each mode
is converted to a classical perturbation. The perturbation is gaussian 
so that the only connected  correlator is 

\bea
\vev{A_{\lambda\bfk}A^*_{\lambda'\bfkp}} &=& (2\pi)^3 \delta_{\lambda\lambda'}
\delta(\bfk-\bfk') \frac{2\pi^2}{k^3} \calp_\lambda(t,k) \\
\frac{2\pi^2}{k^3} \calp_\lambda(t,k) &=& |A_{\lambda k}(t)|^2
. \dlabel{hata} \eea
The spectrum $\calp_\lambda(t,k)$ determines the expectation value 
of $A^2_\lambda(t,\bfx)$ which is also its spatial
average:
\be
\vev{ A^2_\lambda (t)}  = \int^\infty_0 \frac{dk}k \calp_\lambda(t,k)
. \dlabel{asquared} \ee

The vacuum fluctuation of the physical  transverse mode  
$B_\bot=A_\bot/a$ is the same as that of 
a scalar field with mass-squared $m^2$. Well before horizon exit the action is that
of a harmonic oscillator:
\be
S_\bot = (2\pi)\mthree \int d\eta d^3k \frac12 \(|\pa_\eta A_{\bot\bfk}|^2 -
 k^2  |A_{\bot\bfk}|^2 \).
\ee
We choose the mode function
\be
A_{\bot k}(\eta)  = e^{-ik\eta}/\sqrt{2k}, 
\dlabel{perpmode} \ee
so that $\hat A_{\bot\bfk}$ describes massless particles with momentum and energy
$p=k/a$, and we choose the vacuum state. This choice is practically mandatory,
because an occupation number $n_k\gsim 1$ on cosmological scales
would almost certainly generate too much
positive pressure. To be precise, this pressure would exceed the inflationary
pressure $-3\mpl^2H^2$ at the beginning of inflation, unless fewer than
$\ln(\mpl/H)/2$ $e$-folds of inflation occur before the observable universe leaves the 
horizon \cite{treview,abook}.

The evolution equation for $\aperk$ is
\be
\pa_\eta^2 \aperk  + \( k^2 + a^2M^2 \) \aperk = 0.
\dlabel{tranevolution} \ee
Well after horizon exit its solution has  constant phase which means that 
we have a classical field $A_{\bot\bfk}$. (To be precise, a measurement could create
such a field.)  Just a few Hubble times after
horizon exit we can take $m^2=0$ and $H$ constant. Then 
the  spectrum of the physical field is 
\be
\calp_{B_\bot}(k) = \frac{k^3}{2\pi^2} \frac {\left|A_{\bot k}\right|^2}{a^2} = \left( \frac H{2\pi} \right)^2. 
\ee
Since the spatial gradient is now negligible, the
subsequent evolution is given by \eq{unpert}.
Taking $H$ to be constant,  and assuming also $m\ll H$,  \eq{unpert} gives 
$B_i\propto t^{-m^2/3H}$.  Taking also $\dot H/H^2$ to be constant,
this gives 
the  spectrum evaluated at a time when all relevant scales have left the 
horizon:\footnote
{The approximations $H$ and $\dot H/H^2$ constant should be adequate during
the 15 or so $e$-folds that occur while cosmological scales are leaving 
the horizon.}
 \be
\calp_\bot(t,k) = \( \frac H{2\pi}\)^2 \( \frac k {aH} \)^{2\eta-2\epsilon},\qquad
\eta\equiv m^2/3H^2,\quad \epsilon \equiv -\dot H/H^2
, \dlabel{transpec} \ee
with $\epsilon$ and $\eta$ constant.
(Note that this $\eta$ has nothing to do with the conformal time.)

The action for the longitudinal mode is \cite{peloso1}
\be
S_{||}= \left(2\pi\right)^{-3}\int\mathrm{d}\eta\,\mathrm{d}^{3}k\frac{1}{2}M^{2}
\left[\frac
{ \left|\pa_\eta\alon\right|^{2} }{ p^2+M^{2} }- a^{2}\left|\alon\right|^{2} \right].
\label{eq:Action-long} \ee 
As long as $m\ll H$, we see that  in the sub-horizon regime where $p$ is slowly varying  $\left. \tilde A_{\lon\bfk} \equiv \sqrt2 \alon H/p\right.$ has the  action

\be
S_\lon \simeq  -(2\pi)\mthree \int d\eta d^3k \frac12 \(|\pa_\eta \tilde A_{\lon\bfk}|^2 -
 k^2  |\tilde A_{\lon\bfk}|^2 \).
 \ee
Except for the minus sign this is the same as for the transverse mode. We choose
the same mode function for $\tilde A_{\lon k}$ as for $A_{\bot k}$ in \eq{perpmode}, corresponding to

\be
\alonk = \frac p{\sqrt 2 H}  \frac 1{\sqrt {2 k}} e^{-ik\eta}, 
\dlabel{lonmode} \ee
and we choose  the vacuum state. It was presumed in \cite{dklr} that $\hat A_{\lon\bfk}$ 
describes particles with energy $-(k/a)$ and we verify that in the next section.
Then the vacuum choice is practically mandatory, because the argument given for
the transverse mode holds, except that we deal with negative energy density and pressure,
and it is the energy density not the pressure which would spoil inflation \cite{dklr}.

The evolution equation for $\alonk$ is \cite{peloso2}
\begin{equation}
\left[\partial_{\eta}^{2}+aH\frac{2p^{2}}{p^{2}+M^{2}}
\( 1 +\frac1{2HM^2}\frac{d(M^2)}{dt} \) \pa_\eta
+a^{2}\left(p^{2}+M^{2}\right)
\right]\alonk=0.\dlabel{lonevolution}
\end{equation}
 The  second term in the first round bracket is negligible during inflation, because then it is equal to $-\dot H/H^2$. If our action continues to hold after inflation, this term may become significant and if $m$ is nonzero $M^2$ will rise through zero
making this term singular.

During inflation with $|m^2|\ll H^2$ the term  $(p^2+M^2)$ passes through zero around the time of horizon exit, but $\alonk$  is regular there \cite{dklr,peloso2} and so is \cite{peloso2} $\dalonk/(p^2+M^2)$. Taking $m^2=0$ and $H$ constant, one finds soon after horizon exit the spectrum $\calp_{||} = 2 \calp_\bot$. The subsequent  evolution $\alonk$ is  again given by \eq{unpert} so that when cosmological scales have left the horizon

\be
\calp_\lon(t,k) = 2\calp_\bot(t,k). 
\dlabel{spectra}  \ee
with $\calp_\bot$ given by \eq{transpec}. 

After inflation is over, $\delta A_\mu$ can generate a contribution to the curvature perturbation through the curvaton mechanism \cite{curvaton} or one of its generalisations \cite{newbook}. Through the  $\delta N$ formula \cite{deltan,nonlindeltan,sy,dklr}, \eqs{transpec}{spectra}  then give  the  spectrum,  bispectrum  etc.\ of this
 contribution. Because of the factor 2 in \eq{spectra}, they exhibit distinctive anisotropy \cite{dklr,more1}.

On the assumption that $A_\mu$ instead describes the electromagnetic field \cite{turnerwidrow}, the physical vector potential $B_i$ becomes time-independent in the super-horizon regime, and its transverse component gives a time-independent primordial magnetic field ${\bf B}\sub{mag}= \,\mbox{curl}\, {\bf B}$. The flat spectrum of $\aper$ makes the spectrum of ${\bf B}\sub{mag}$ go like $p^2$, which  makes it perhaps difficult to generate a useful primordial magnetic field on cosmological scales, and one  might prefer the spectrum of $\aper$ to go like $p\mtwo$ \cite{my} so that  ${\bf B}\sub{mag}$ has a flat spectrum. The longitudinal mode of the electromagnetic field in this scenario has not been mentioned in the literature.

\section{Energy density: general expression and vacuum fluctuation}

\dlabel{sthree}

\subsection{General expression}

The interaction $RA^2/6$  means that we are not dealing with Einstein gravity, but
we  still  define the energy-momentum tensor through the Einstein equation,
\be
R_\mn - \frac12 g_\mn R = \mpl\mtwo T_\mn . 
\ee
The contribution of bosonic fields is

\begin{equation}
T_{\mu\nu}=\frac{2}{\sqrt{-g}}\frac{\delta\left(\sqrt{-g}\mathcal{L}\right)}
{\delta g^{\mu\nu}}.
\dlabel{tmn} \end{equation}
Since we used the Robertson-Walker metric to deal with $A_\mu$, we do the same to evaluate
the contribution of $A_\mu$ to \eq{tmn}. It is \cite{vecinf}

\begin{equation}
T_{\mu}^{\nu}=T_{1\mu}^{\,\nu}+T_{2\mu}^{\,\nu},
\end{equation}
where $T_{1\mu}^{\,\nu}$ is the part of the energy-momentum tensor
which has the same form as the minimally coupled vector field

\begin{equation}
T_{1\mu}^{\,\nu}=\frac{1}{4}\delta_{\mu}^{\nu}F^{2}-F_{\mu\kappa}F^{\nu\kappa}-\frac{1}{2}M^{2}\left(\delta_{\mu}^{\nu}A^{2}-2A_{\mu}A^{\nu}\right),
\label{eq:T1mn}\end{equation}
while $T_{2\mu}^{\,\nu}$ is an additional term due to non-minimal coupling to gravity 

\begin{equation}
T_{2\mu}^{\,\nu}=\frac{1}{6}\left(R_{\mu}^{\nu}+\delta_{\mu}^{\nu}\nabla_{\kappa}\nabla^{\kappa}-\nabla_{\mu}\nabla^{\nu}\right)A^{2}.
\label{eq:T2mn}\end{equation}
The latter will become negligible at late times, when $R$ becomes negligible.

The contribution of the unperturbed physical field to the energy density and pressure are
the same as for a scalar field \cite{vecinf}:

\be
\rho= \frac12 \( \dot B^2  + \frac12 m^2 B^2 \),\qquad P = 
\frac12 \( \dot B^2  - \frac12 m^2 B^2 \)
. \dlabel{unpertrho} \ee
(They satisfy the   continuity equation $\dot\rho=-3H(\rho+P)$ by virtue of the field equation
\eqreff{unpert}.) 
The anisotropic stress is of the same order as the pressure, which when
 inserted into the Einstein equation
would be inconsistent with the assumed Robertson-Walker metric. We therefore
assume that the contribution of the unperturbed vector field to $T_\mn$ is negligible.
(Instead, one can invoke many randomly oriented vector 
fields, to arrive at a vector inflation
model \cite{vecinf,vecinf2}.)

\subsection{Vacuum fluctuation}

Now we consider  for the first time the   contribution of $\delta A$ to $T_\mn$.
We will take  the spatial average which kills the anisotropic stress and gives  $\rho$
and $P$ as a mode sum. For $\rho$ we write
\be
\rho(t) = \int^\infty_0 \frac {dk}k \calp_\rho(t,k),\qquad \calp_\rho=2\calp_{\rho\bot}
+ \calp_{\rho\lon}
, \dlabel{rhooft} \ee
where the superscripts indicate transverse and longitudinal contributions to the spectrum.
Then we write
\be
\calp_{\rho\bot} = \calp_{\rho1\bot} +  \calp_{\rho2\bot}
, \ee
and similarly for $\calp_{\rho\lon}$, 
where the first term comes from \eq{eq:T1mn} and the second comes from \eq{eq:T2mn},
and we treat the pressure $P$ in the same way.

In this section we consider the contributions generated by the vacuum fluctuation.
The  transverse contributions to the energy density are

\begin{eqnarray}
\calp_{\rho1\bot} & = & \frac{ap^{3}}{\left(2\pi\right)^{2}}\left[\left| 
\daperk\right|^{2}+\left(p^{2}+M^{2}\right)\left|\aperk\right|^{2}\right], \dlabel{tran1}\\
\calp_{\rho2\bot} & = & \frac{ap^{3}}{\left(2\pi\right)^{2}}
\left[\left(3H^{2}+\dot{H}\right)\left|\aperk\right|^{2}-
H\left(\daperk\aperk^{*}+\aperk \daperk^{*}\right)\right]. \dlabel{tran2}
\end{eqnarray}
And the longitudinal contributions are

\begin{eqnarray}
\calp_{\rho1\lon} & = & \frac{ap^{3}}{\left(2\pi\right)^{2}}\left[
\frac{M^{2}}{p^{2}+M^{2}}\left|{\dalonk}\right|^{2}+M^{2}\left|\alonk\right|^{2}\right],
\dlabel{long1}\\
\calp_{\rho2\lon} & = & \frac{ap^{3}}{\left(2\pi\right)^{2}}\left[
-\frac{\left(7H^{2}+\dot{H}+\frac{2H}{M^2}\frac{ d(M)^2}{dt} \)
p^{2}}{\left(p^{2}+M^{2}\right)^{2}}
\left|{\dalonk}\right|^{2}+\left(3H^{2}+\dot{H}\right)\left|\alonk\right|^{2}\right.
\nonumber \\
 &  & \left.-H\frac{2p^{2}+M^{2}}{p^{2}+M^{2}}
\left({\dalonk}\alonk^{*}+\alonk{\dalonk}^{*}\right)\right]. \dlabel{long2}
 \end{eqnarray}
 The contributions to the pressure are  

\begin{eqnarray}
\calp_{P1\bot} & = & \frac{ap^{3}}{12\pi^{2}}\left[\left|\daperk\right|^{2}
+\left(p^{2}-M^{2}\right)\left|\aperk\right|^{2}\right],\\
\calp_{P2\bot} & = & \frac{ap^{3}}{12\pi^{2}}\left[2\left|\daperk\right|^{2}
-3\left(H^{2}+\dot{H}\right)\left|\aperk\right|^{2}-\right.\nonumber \\
 &  & \left.-2H\left(\daperk\aperk^{*}+\aperk \daperk^{*}\right)
+\left( \ddaperk\aperk^{*}+\aperk \ddaperk^{*}\right)\right].
\end{eqnarray}
and

\bea
\calp_{P1\lon} & = & \frac{ap^{3}}{12\pi^{2}}
\left[\frac{M^{2}\left(3p^{2}M^{2}+M^{2}\right)}{\left(p^{2}+M^{2}\right)^{2}}\left|{\dalonk}\right|^{2}-M^{2}\left|\alonk\right|^{2}\right],\\
\calp_{P2\lon} & = & -\frac{ap^{3}}{12\pi^{2}}
\left[\left(\frac{p^{2}L^{2}}{\left(p^{2}+M^{2}\right)^{2}}
-2\frac{2p^{2}+M^{2}}{p^{2}+M^{2}}\right)\left|{\dalonk}\right|^{2}
+\left(3H^{2}+3\dot{H}+2p^{2}\right)\left|\alonk\right|^{2}\right.\nonumber \\
 &+& \left.\frac{p^{2}\left(11H+\frac3{M^2} \frac{d M^2}{dt} 
\right)+2HM^{2}}{p^{2}+M^{2}}\left({\dalonk}\alonk^{*}+\alonk{\dalonk}^{*}\right)
-\left({\ddalonk}\alonk^{*}+\alonk{\ddalonk}^{*}\right)\right],\qquad
\end{eqnarray}
where

\be
L^2\equiv 21H^2 - 7\dot H  + 20 \frac H{M^2}  \frac{d (M)^2}{dt}
+\frac 6{M^4} \(  \frac{d (M)^2}{dt} \)^2 - \frac 2{M^2}\frac{d^2 (M^2)}{dt^2}
. \ee
By virtue of the \eqs{tranevolution}{lonevolution}, the 
 continuity equation is satisfied by each mode separately. 
These expressions remain  finite when  $p^2+M^2$ goes through zero around the time of
horizon exit, because $\dalonk/(p^2+M^2)$ remains finite.
As discussed in the Introduction, we ignore the singularity that will  occur after inflation
if our action remains valid with nonzero $m$ so that $M^2$ rises through zero.

Consider first the super-horizon regime  $p^2\ll H^2$, in which the
vacuum fluctuation has generated a classical perturbation.
Ignoring the possible epoch
when $M^2$ rises through  zero this is also the regime $p^2\ll |M^2|$. If also
$p\ll m$  the spatial gradient
of $\delta B_i$ is negligible and  we can set
 $p=0$ in \eqst{tran1}{long2}. Putting the result into \eq{rhooft}, we arrive
at the classical quantity $\rho_{\delta B}$. Using \eq{asquared} and the analogous 
expression for $\vev{|\delta \dot B|^2}$, it  can be written

\be
\rho_{\delta B} = 
\frac12\( \vev{|\dot {\delta B}|^2} + m^2 \vev{| \delta B|^2} \). 
\ee
This is of the same form as \eq{unpertrho},  and it holds at each position because the spatial gradient
is negligible. The part of $\delta B$ that comes from scales much bigger than the observable universe cannot be distinguished from the unperturbed quantity, and neither can its energy density.

If instead $m^2=0$ (or least negligible compared with $p^2$) we find to leading order in $p^2$
\be
\calp_{\rho\bot} = \frac{ap^3}{(2\pi)^2} p^2 
|\aperk|^2,\qquad  \calp_{\rho\lon}=0
. \dlabel{mageq} \ee
Using \eqs{hata}{rhooft}, this  gives the known  result
$\rho_{\delta B} = \vev{B^2\sub{mag}/2}$ where 
${\bf B}\sub{mag}= \,\mbox{curl}\, {\bf B}$ is the magnetic field.
We see that the longitudinal mode, ignored in previous work,
 gives in fact no contribution to $\rho$.

Now consider the regime $m\ll H$ and $p\gg H$. The latter condition (sub-horizon)
means that  $\hat A_{\bot\bfk}$ 
and $\hat A_{\lon\bfk}$ describe practically massless
particles, and using \eqs{perpmode}{lonmode} and \eqst{tran1}{long2} we find
for the vacuum state
\be
\calp\su{vac}_{\rho\bot} = -\calp\su{vac}_{\rho\lon} = \frac {p^4}{4\pi^2}
. \dlabel{rhovacs} \ee
These contributions to the energy density 
diverge in the ultra-violet and as usual we drop them. 

\section{Energy density: particles}

\dlabel{sfour}

Instead of the vacuum state, suppose now that the state corresponds to 
 occupation number $n_k$, which is independent of direction.
 The operator expression 
\eq{hata} and the mathematics of the harmonic oscillator imply that we should then
make the replacement $\calp\su{vac}\to 2n_k \calp\su{vac}$.  Then, remembering that the
density of states is $1/(2\pi)^3$, \eq{rhovacs} shows that each transverse particle
carries energy $p$ while each longitudinal particle carries energy $-p$.
This holds  for as long as $m\ll H$; in other words,
 unless and until we encounter the singularity epoch when $M^2$ rises through
zero.

The negative energy density of the longitudinal mode occurs when the 
 kinetic term in \eq{eq:Action-long} is negative, which is caused by
 the non-minimal coupling to gravity. Since the longitudinal particles
carry negative energy, energy conservation allows them to be created from the vacuum
along with ordinary  particles (and/or transverse particles) that carry positive energy.

A scalar field $\phi$
living in flat spacetime (minimal coupling to gravity),
whose kinetic term is $-1$ times the canonical one, is called a ghost.
Using flat spacetime quantum field theory, one can estimate the rate of processes like
\be
\mbox{vacuum }\rightarrow \phi + \phi + \gamma + \gamma
. \ee
It is found \cite{ghost} that the rate for this process would violate observational constraints
on either air showers (assuming Lorentz invariance) or would require an implausibly
low ultra-violet cutoff on the effective field theory (allowing a violation of Lorentz
invariance). 

The situation for our longitudinal mode is quite different. The non-minimal coupling
to gravity, as well the condition  $m\ll H$,  mean 
that the  creation rate cannot be calculated using flat spacetime theory.
Also, unless $m$ is strictly zero, the creation occurs only in the early universe
which means that it is not constrained by direct observation. 

After some particles have been created, their energy density redshifts like $1/a^4$ while they are relativistic, and like $1/a^3$ after they have become non-relativistic. The epoch of transition between these two regimes will be different for the longitudinal particles and the ordinary particles, which means that the initial cancellation between their energy densities will not be preserved. Much as in the warm inflation scenario, the energy density of the created particles will come into equilibrium, where the rate of creation balances the redshift. In order not to affect the creation of longitudinal perturbations from the vacuum, the longitudinal and transverse occupation numbers should be $n_k\ll 1$. (The presence of ordinary particles, giving zero total energy density, of course invalidates  the argument given earlier for $n_k \ll 1$ being mandatory.)

It is not clear how to estimate the creation rate, which as already stated has nothing to do with flat spacetime field theory. Since the creation is caused by the coupling $R A^2/6$ though, significant creation will not persist until very late times even if $m^2$ remains negligible (so that the energy density of the longitudinal particles remains negative). This is because $R$ becomes negligible at late times compared with all other relevant energy scales. In  particular,  \eq{eq:Ricci-scl} gives  at the present epoch $|R|\sim  H_0^2 \sim 10^{-66}\eV^2$. This means that longitudinal photon creation does not necessarily invalidate the use of our action as a mechanism for generating a primordial magnetic field, even though it implies in principle the existence of longitudinal photons.

\section{Avoiding a mass term}
\dlabel{sfive}

After inflation, $R$ given by \eq{eq:Ricci-scl} 
remains negative and at most of order $H^2$.\footnote
{With strict radiation domination $R=0$, but non-Abelian interactions
are expected to make $R$ a significant fraction of $H^2$ even then
\cite{peloso2}.} A significant  mass term will therefore cause
$M^2$ to rise through zero at some point, leading to the singularities 
mentioned earlier. In this section we consider two ways in which such a term might be 
avoided.

The most promising way is to set $m=0$, but invoke a coupling $g^2\phi^2 A^2$
or $g^2 |\phi|^2 A^2$, to some real or complex  scalar field. 
The latter case, with the coupling coming from a gauge coupling,  is particularly attractive,
among other things because it justifies the canonical kinetic term.
(Of course the gauge symmetry is broken by the $RA^2/6$ coupling, but that term disappears
in the flat spacetime limit that will be an excellent approximation at late times.)\footnote
{The view we are taking in this paper, that a gauge symmetry broken only by the $RA^2/6$ coupling
may be relevant, is more up-beat than the one taken in Section 7.1 of \cite{dklr}.}
If $\phi$ has zero vev, such a coupling would allow the perturbation $\delta A$ to affect
the mass of $\phi$, offering the possibility of 
an inhomogeneous decay mechanism for generating the curvature perturbation.
Alternatively, if $\phi$ has nonzero vev this coupling could generate an effective
mass for the vector field,
 which may avoid the singularity occurring when $A_\mu$
has  a mass term and allow say a curvaton mechanism.

Another  possibility  
might be to drop the coupling $RA^2/6$, and to introduce a
negative mass-squared $m^2\simeq -2H^2$ during inflation. To stabilize $A_\mu$
one would then have to introduce a vector field potential
$V(A^2)$ giving $A^2$ a vev at which $V(A^2)=0$, which if renormalizable
would be of the form $V(A^2) = - m^2 A^2 + \lambda A^4$. 
This possibility (without an explicit form for $V(A^2)$)
was invoked \cite{kostas} to provide a curvaton mechanism. It might though turn out,
when the field equation and energy density are worked out, that a singularity
still develops when the effective mass-squared $d V/d (A^2)$ of the perturbation
passes through zero.

Finally, we point out that instead of a single field  $A$ we might be dealing with a non-Abelian  multiplet. Indeed, if  $A$ describes the electromagnetic field that is certainly the case, because $A_\mu$  that field is part of at least the electroweak $SU(2)\times U(1)$ multiplet which in turn may be  part of a  GUT multiplet. If $A$ is a multiplet with (say) the $RA^2/6$ coupling to gravity, the issues we have raised will need to be revisited which has not been done. All that has  been done in this direction (without reference to a specific mechanism for generating $\delta A$) is a calculation of the non-gaussianity of $\delta A$ soon after horizon exit, that is generated by its self-coupling \cite{nonabelian}. However, the self-interaction cannot be too strong, or the non-linear evolution of $A_\mu$ would not be viable. This places an upper bound on the gauge coupling, which is presently unknown and which might be violated by the Standard Model running couplings.

\section{Conclusion}

\dlabel{ssix}

A light vector field with the canonical kinetic term and minimal coupling to gravity
is invariant under conformal transformations, and as a result it
cannot contribute significantly to the primordial curvature perturbation (at least on
cosmological scales). Nor can such a field (identified in that case as the electromagnetic
vector potential) generate a viable primordial magnetic field.\footnote
{In fact, due to the conformal invariance there is no classical magnetic field
at all on  scales that have entered the horizon, because there is no
 Bogoliubov transformation.}

To avoid this situation one must break the conformal invariance, by
introducing  a non-minimal kinetic term
and/or a non-minimal coupling to gravity. We have explored the latter possibility,
adopting the simplest viable coupling which is $RA^2/6$. That the  factor $1/6$
works was discovered by accident \cite{turnerwidrow}, and it is not known whether
the fact that it works is related at some deep level to the fact that a 
 coupling $R\phi^2/6$ of a {\em scalar} field {\em restores} the otherwise broken 
conformal invariance.

In this paper we have addressed the issues, raised in \cite{peloso1,peloso2},
that call into question the health of this coupling with regard to the 
 longitudinal perturbation. To facilitate this, we have for the first time
calculated the contribution of the vector field perturbation to the spatially-averaged
energy density of the universe. We have verified that the contribution of the
longitudinal component is negative unless and until the mass
of the vector field becomes of order the Hubble parameter.
Then  longitudinal particles can at some level be created from the vacuum
along with ordinary particles,
but we have noted that such creation will become negligible at late times
because the coupling of the field to gravity will be negligible.
 In this
regard, the longitudinal field is quite different from a scalar field
which simply has the wrong-sign kinetic term; such a field is called a ghost and
is indeed forbidden because it would lead to the creation of too many photons
from the present-day vacuum \cite{ghost}.\footnote
{This difference is not recognised in  footnote 16 of \cite{peloso2},
which states that the mass zero case is ruled out  by \cite{ghost}.}

We have also considered the other issue, which is that the linear evolution becomes
singular at certain epochs if the action of the
vector field has a nonzero mass term.  We have pointed
out that while this signals a failure of the linear theory, it remains to be seen
whether the full theory is sick. In addition, we have mentioned ways in which 
a nonzero mass term can be replaced by something more complicated, which may well
avoid the singularities.

\section{Acknowledgements}
DHL and MK are supported by EU grant MRTN-CT-2006-035863 and DHL by EU grant UNILHC23792. MK also acknowledges
 the support of Lancaster University's Department of Physics.


\begin{thebibliography}{99}

\bibitem{kostas}
  K.~Dimopoulos,
``Can a vector field be responsible for the curvature perturbation in the
universe?,''
  Phys.\ Rev.\  D {\bf 74} (2006) 083502.

\bibitem{sy}
  S.~Yokoyama and J.~Soda,
``Primordial statistical anisotropy generated at the end of inflation,''
  JCAP {\bf 0808} (2008) 005.

\bibitem{KDandMK}
  K.~Dimopoulos and M.~Kar\v{c}iauskas,
``Non-minimally coupled vector curvaton,''
  JHEP {\bf 0807} (2008) 119;
  
\bibitem{dklr}
  K.~Dimopoulos, M.~Kar\v{c}iauskas, D.~H.~Lyth and Y.~Rodriguez,
  ``Statistical anisotropy of the curvature perturbation from vector field
  perturbations,''
  JCAP {\bf 0905} (2009) 013

\bibitem{more1}
  M.~Kar\v{c}iauskas, K.~Dimopoulos and D.~H.~Lyth,
``Anisotropic non-Gaussianity from vector field perturbations,''
  Phys.\ Rev.\  D {\bf 80} (2009) 023509;
  C.~A.~Valenzuela-Toledo, Y.~Rodriguez and D.~H.~Lyth,
``Non-gaussianity at tree- and one-loop levels from vector field
perturbations,''
  Phys.\ Rev.\  D {\bf 80} (2009) 103519;
  C.~A.~Valenzuela-Toledo and Y.~Rodriguez,
``Non-gaussianity from the trispectrum and vector field perturbations,''
  Phys.\ Lett.\  B {\bf 685} (2010) 120.
  
\bibitem{more2}
  K.~Dimopoulos, M.~Kar\v{c}iauskas and J.~M.~Wagstaff,
``Vector Curvaton with varying Kinetic Function,''
  Phys.\ Rev.\  D {\bf 81} (2010) 023522;
 K.~Dimopoulos, M.~Kar\v{c}iauskas and J.~M.~Wagstaff,
 ``Vector Curvaton without Instabilities,''
  Phys.\ Lett.\  B {\bf 683} (2010) 298;

\bibitem{nonabelian}
  N.~Bartolo, E.~Dimastrogiovanni, S.~Matarrese and A.~Riotto,
  ``Anisotropic bispectrum of curvature perturbations from primordial
  non-Abelian vector fields,''
  JCAP {\bf 0910} (2009) 015;
  N.~Bartolo, E.~Dimastrogiovanni, S.~Matarrese and A.~Riotto,
``Anisotropic Trispectrum of Curvature Perturbations Induced by Primordial
Non-Abelian Vector Fields,''
  JCAP {\bf 0911} (2009) 028;
  E.~Dimastrogiovanni, N.~Bartolo, S.~Matarrese and A.~Riotto,
``Non-Gaussianity and statistical anisotropy from vector field populated
inflationary models,''
  arXiv:1001.4049 [astro-ph.CO].

\bibitem{vecinf}
  A.~Golovnev, V.~Mukhanov and V.~Vanchurin,
``Vector Inflation,''
  JCAP {\bf 0806} (2008) 009;

\bibitem{vecinf2}
  A.~Golovnev, V.~Mukhanov and V.~Vanchurin,
 ``Gravitational waves in vector inflation,''
  JCAP {\bf 0811} (2008) 018;
  A.~Golovnev and V.~Vanchurin,
 ``Cosmological perturbations from vector inflation,''
  Phys.\ Rev.\  D {\bf 79} (2009) 103524.

\bibitem{peloso1}
 B.~Himmetoglu, C.~R.~Contaldi and M.~Peloso,
  ``Instability of anisotropic cosmological solutions supported by vector
  fields,''
  Phys.\ Rev.\ Lett.\  {\bf 102} (2009) 111301;
 B.~Himmetoglu, C.~R.~Contaldi and M.~Peloso,
 ``Instability of the ACW model, and problems with massive vectors during
 inflation,''
  Phys.\ Rev.\  D {\bf 79} (2009) 063517.

\bibitem{peloso2}
 B.~Himmetoglu, C.~R.~Contaldi and M.~Peloso,
``Ghost instabilities of cosmological models with vector fields nonminimally
coupled to the curvature,''
  Phys.\ Rev.\  D {\bf 80} (2009) 123530.

\bibitem{turnerwidrow}
  M.~S.~Turner and L.~M.~Widrow,
``Inflation Produced, Large Scale Magnetic Fields,''
  Phys.\ Rev.\  D {\bf 37} (1988) 2743.

\bibitem{curvaton}
  A.~D.~Linde and V.~F.~Mukhanov,
``Nongaussian isocurvature perturbations from inflation,''
  Phys.\ Rev.\  D {\bf 56} (1997) 535;
  D.~H.~Lyth and D.~Wands,
``Generating the curvature perturbation without an inflaton,''
  Phys.\ Lett.\  B {\bf 524} (2002) 5;
  T.~Moroi and T.~Takahashi,
``Effects of cosmological moduli fields on cosmic microwave background,''
  Phys.\ Lett.\  B {\bf 522} (2001) 215
  [Erratum-ibid.\  B {\bf 539} (2002) 303];
D. H. Lyth, C. Ungarelli, and D. Wands,
  ``The primordial density perturbation in the curvaton scenario,''                    
  Phys. Rev.  D {\bf 67}, 023503 (2003).

\bibitem{newbook}
 D. H. Lyth and A. R. Liddle, {\it The Primordial Density Perturbation                  
(Cosmology, Inflation and the origin of Structure) },
Cambridge University Press, 2009.

\bibitem{treview}
 D.~H.~Lyth and A.~Riotto,
 ``Particle physics models of inflation and the cosmological density
  perturbation,''
  Phys.\ Rept.\  {\bf 314} (1999) 1

\bibitem{abook}
A. R. Liddle and D. H. Lyth,
{\it Cosmological Inflation and Large-Scale Structure},
Cambridge University Press, 2000.

\bibitem{deltan}
  A.~A.~Starobinsky,
``Multicomponent De Sitter (Inflationary) Stages And The Generation Of
Perturbations,''
   JETP Lett.\  {\bf 42}, 152 (1985)
   [Pisma Zh.\ Eksp.\ Teor.\ Fiz.\  {\bf 42}, 124 (1985)];
  M.~Sasaki   and E.~D.~Stewart,
``A General analytic formula for the spectral index of the density
perturbations produced during inflation,''
   Prog.\ Theor.\ Phys.\  {\bf 95} (1996) 71;

\bibitem{nonlindeltan}
D.~H.~Lyth, K.~A.~Malik and M.~Sasaki,
``A general proof of the conservation of the curvature perturbation,''
JCAP {\bf 0505}, 004 (2005);
  D.~H.~Lyth and Y.~Rodriguez,
``The inflationary prediction for primordial non-gaussianity,''
   Phys.\ Rev.\ Lett.\  {\bf 95} (2005) 121302.

\bibitem{my}
  J.~Martin and J.~Yokoyama,
``Generation of Large-Scale Magnetic Fields in Single-Field Inflation,''
  JCAP {\bf 0801} (2008) 025.

\bibitem{ghost}
J.~M.~Cline, S.~Jeon and G.~D.~Moore,
``The phantom menaced: Constraints on low-energy effective ghosts,''
  Phys.\ Rev.\  D {\bf 70} (2004) 043543.

\end{thebibliography}
\end{document}